# Determining Phonon Mean Free Path Spectrum by Ballistic Phonon Resistance within a Nanoslot-Patterned Thin Film

Qing Hao[*], Yue Xiao, and Qiyu Chen

Department of Aerospace and Mechanical Engineering, University of Arizona

Tucson, AZ, USA 85721-0119

Electronic mail: qinghao@email.arizona.edu

**Abstract**

At micro- to nano-scales, classical size effects in heat conduction play an important role in suppressing the thermal transport process. Such effects occur when the characteristic lengths become commensurate to the mean free paths (MFPs) of heat carriers that are mainly phonons for nonmetallic crystals. Beyond existing experimental efforts on thin films using laser-induced thermal gratings, this work provides the complete theoretical analysis for a new approach to extract the effective phonon MFP distribution for the in-plane heat conduction within a thin film or flake-like sample. In this approach, nanoslots are patterned on a suspended thin film. Phonons will transport ballistically through the neck region between adjacent nanoslots if the phonon MFPs are much longer than the neck width. The associated "ballistic thermal resistance" for varied neck dimensions can then be used to reconstruct the phonon MFP distribution within the film. The technique can be further extended to two-dimensional materials when the relaxation time approximation is reasonably accurate.

## 1. INTRODUCTION

One central topic of nanoscale heat transfer is to tailor the thermal properties of a material with the nanostructuring approach. When the structure size becomes comparable with or shorter than the bulk mean free path (MFP) $\Lambda$ of some phonons, these phonons are forced to scatter more at boundaries or interfaces and the thermal transport can be strongly suppressed.[1,2] In principle, phonon MFPs in a bulk material usually have a wide distribution, ranging from nanometers to millimeters.[3,4] As one major input for phonon transport analysis, the employed phonon MFPs can largely affect the calculation results.

In simplified analysis, the gray-body approximation is applied to a bulk material, in which all phonons share the same phonon MFP and group velocity. Following this, an averaged $\Lambda$~43 nm is estimated for bulk Si using the kinetic relationship, $k_L = cv_g\Lambda/3$, where $k_L$, $c$ and $v_g$ are the lattice thermal conductivity, volumetric phonon specific heat, and averaged phonon group velocity, respectively.[5] This is in contrast with thin-film measurements that suggest the dominant phonon MFP in Si to be ~300 nm at 300 K.[6] More recent first-principles calculations for energy-dependent $\Lambda$ suggest that around 50% of the room-temperature $k_L$ is contributed by phonons with MFPs longer than 1 µm.[7] Correspondingly, a constant $\Lambda$~43 nm fails to explain the remarkable phonon size effects observed in microporous Si films,[8] which can only be explained after considering $\Lambda$ dependence on the phonon angular frequency $\omega$ and branch $i$.[9]

Tremendous computational efforts have been dedicated to extracting the phonon MFP distribution. In the simplest approach, the temperature-dependent $k_L$ of a bulk material can be fitted to determine the parameters used in the $\Lambda(\omega, i)$ expressions for different phonon scattering mechanisms.[10] Even for the same phonon scattering mechanism, however, different $\Lambda(\omega, i)$ expressions can be found in the literature. Large uncertainties exist due to employed $\Lambda(\omega, i)$ expressions and numerous fitting parameters. Without curve fittings, $\Lambda(\omega, i)$ can now be directly computed from first principles[7, 11, 12] and can be consistent with the phonon MFP spectrum obtained in pump probe measurements.[13, 14] Beyond Si, first-principles computations of phonon MFPs have also been carried out on other materials with relatively simple atomic structures.[3, 4] For complicated materials with a large number of atoms per primitive cell,[15, 16] such computations become very challenging due to the huge computational load. In this aspect, experimental techniques to extract the phonon MFP distribution in arbitrary materials is important not only for validating computations but also for novel materials that are difficult to be studied by first-principles computations.

In experiments, the most successful technique to extract the phonon MFP spectrum is based on pump probe measurements.[1] These measurements use a probe laser to detect the temperature variation of the sample surface that is thermally excited by a pump laser. When the size $D$ of the heating patterns is reduced, the lattice thermal conductivity contributed by a phonon with MFP $\Lambda(\omega, i) \gg D$ scales down, e.g., following $k_L(\omega, i)/k_{L,Bulk}(\omega, i) \sim D/\Lambda(\omega, i)$.[17] The ballistic transport of these long-MFP phonons leads to a ballistic thermal resistance ($R_b$) that is determined by $D$ and $\Lambda(\omega, i)$. The adjusted $D$ includes the thermal penetration depth by changing the modulation frequency,[14, 18, 19] the laser beam diameter,[20] the period of laser-induced thermal gratings on a suspended Si film,[21, 22] width of periodic metal lines,[23, 24] and the size of periodic laser-heated Al nanopillars.[13] Among these, the metallic patterns can cover a wide range of phonon MFPs, whereas other heating patterns cannot achieve sub-200 nm $D$ values. As the most advanced method, a metal-polymer bilayer can be further coated in regions surrounding the Al pillars to prevent laser heating of the sample in these regions and minimize the contribution of these surrounding regions on the reflected probe-beam signal.[13]

Despite the advancement of pump probe techniques, the employed metallic patterns are all fabricated on a wafer-like sample and cannot be easily extended to suspended thin films due to fabrication challenges. On the other hand, adding such metallic patterns may affect the lattice vibration modes within a suspended thin film,[25, 26] which should be avoided for the proposed phonon MFP studies. In recent experiments on Si nanobeams with periodic arrays of Al nanopillars, the reduced $k$ of the nanobeams is also attributed to the surface roughness and the amorphous layer under the pillars.[27] In practice, however, phonon MFP distributions within a thin film are of its own importance when: 1) single-crystal bulk samples cannot be easily synthesized (e.g., $Na_xCoO_{2-\delta}$ micro-flakes as the best $p$-type thermoelectric oxide[28]); 2) thin films are directly grown on a substrate and can be different from their bulk counterpart due to stress/strain and unintentional defects. Therefore, it is critical to extend the phonon MFP studies to thin films and still keep the capability of probing phonons with down to a few nanometer MFPs.

In this work, the complete theoretical analysis is provided for a different technique to experimentally extract the phonon MFP distribution, which is based on the in-plane thermal measurements of nanoslot-patterned thin films. A similar structure has been measured previously but the focus is on the phonon transmission across a row of nanoslots between two adjacent suspended isothermal membranes.[29] In our proposed approach, the focus is instead on the ballistic resistance $R_b$ introduced into a film, as the result of ballistic transport of long-MFP phonons through the neck between patterned nanoslots. In this case, a row of nanoslots should be patterned in the middle of a film that is longer than majority phonon MFPs. By varying the neck width between adjacent nanoslots and thus $R_b$, the phonon MFP distribution within a thin film can be extracted. The phonon MFP distribution for the bulk counterpart can be further derived for special cases, i.e., completely diffusive film-surface phonon reflection at 300 K for a typical film-surface roughness.[30]

In the literature, classical phonon size effects and sometimes phonon wave effects have been widely studied for Si thin films with periodic nanopores.[31-37] However, no studies have been carried out on the phonon MFP reconstruction, mainly due to the difficulty in obtaining a simple but accurate analytical model for the inverse phonon transport analysis. When $k_L$ is computed by modifying the bulk phonon MFPs with a characteristic length or limiting dimension $L_C$ of the nanoporous structure, different $L_C$ expressions[38-43] have been proposed but the predicted $k_L$ for arbitrary nanoporous geometries can often diverge from those given by solving the phonon Boltzmann Transport equation (BTE).[9, 44, 45] In addition, varied pore-edge defects and roughness across a real film can add more complexity to the data analysis.[43, 46-48] In contrast, an accurate characteristic length $L_C$ can be derived for nanoslot-patterned thin films and be directly used for phonon MFP reconstruction in this work. In nanofabrication, patterning a few nanoslots can also have better structure control than patterning numerous periodic nanopores, particularly when a ultra-small $L_C$ is required to probe the $k_L$ contribution of short-MFP phonons.

## 2. EXTRACTING IN-PLANE PHONON MFP DISTRIBUTIONS WITH VARIED BALLISTIC THERMAL RESISTANCES IN FILM-LIKE SAMPLES

### 2.1. Phonon Monte Carlo (MC) simulations for a nanoslot-patterned thin film

Figure 1a shows the studied nanoslot-patterned thin film sandwiched between two thermal reservoirs, which is consistent with typical measurements using two isothermal membranes bridged by the measured thin film.[34, 35] For potential thermoelectric applications, similar structures have also been introduced for graphene to reduce its intrinsically high thermal conductivity.[49] In this work, the structure-dependent thermal conductance of the film is used to extract the phonon MFP distribution from the inverse phonon transport analysis.

In virtual experiments, frequency-dependent phonon MC simulations are carried out to find the $k_L$ of a given nanoslot-patterned thin film.[45] As one method to solve the phonon BTE, phonon MC simulations track the phonon transport and scattering to statistically obtain the BTE solution. To be consistent with aforementioned measurements, the computed thin film is sandwiched between two blackbodies with $\Delta T$ as the applied temperature difference.[45, 50] The temperature profile for a typical 70-nm-thick nanoslot-patterned thin film with a 100 nm neck width is presented in Fig. 1b. For a single period, its left and right sidewalls are treated as specular due to structure symmetry. Other boundaries, such as the film top and bottom surfaces and typically rough nanoslot edges, are assumed to diffusively scatter phonons. This assumption is typically accurate at 300 K or above, where even ~1 nm surface roughness can lead to almost completely diffusive phonon scattering.[30] Experimental studies on the relationship between the surface roughness and the phonon specularity can be found elsewhere.[51-53]

One major problem for phonon MC simulations lies in its poor computational efficiency, which limits the early frequency-dependent phonon MC studies to nanosized structures.[45] This challenge has been recently solved by a new deviational phonon MC technique, which was developed by Péraud and Hadjiconstantinou.[54] Instead of tracking a huge number of phonons, this new technique only focuses on phonons related to the deviation of the phonon distribution function $f$ from the equilibrium $f_0$ (i.e., Bose-Einstein distribution) at a reference temperature. This enables phonon MC simulations for the proposed microsized thin films.

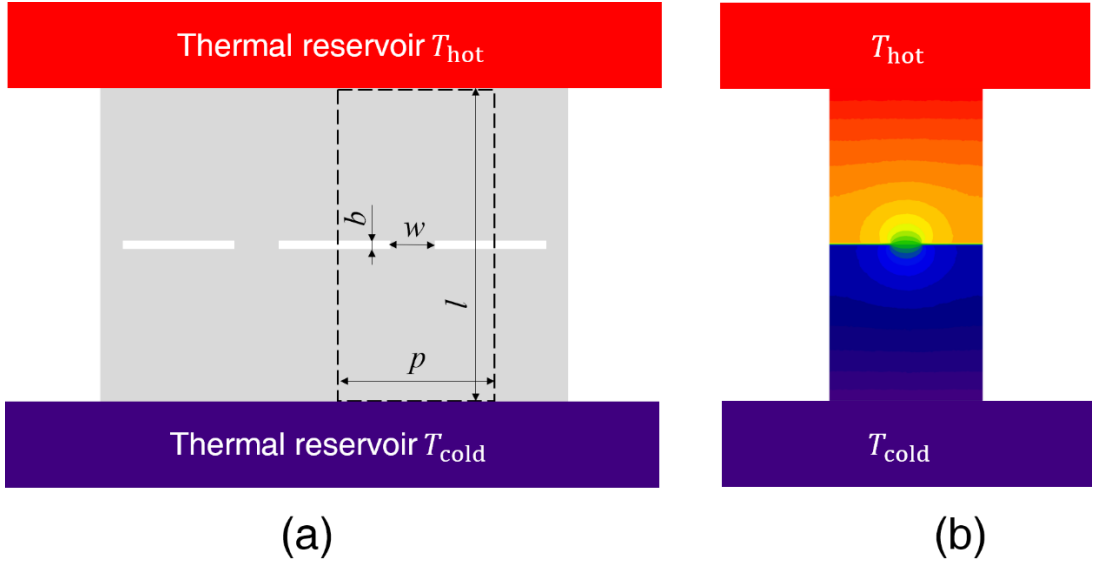

Figure 1. (a) Illustration of a thin film with a row of periodic nanoslots pattered, with pitch $p$, neck width $w$, depth $b$ and film length $l$. A single period is enclosed with the dashed line. (b) A typical temperature profile for one period, as computed by frequency-dependent phonon MC simulations. A temperature jump may further occur at the film-reservoir junctions due to strongly ballistic phonon transport within a film.

## 2.2. Phonon MFP extraction from the simulated thermal conductance of nanoslot-patterned films

The phonon MFP distribution can be constructed by comparing the measured in-plane effective thermal conductivity $k_{R_b}$ of a thin film with patterned nanoslots and that for a reference solid thin film ($k_{Ref}$). In real experiments, both the patterned film and the reference film can be measured to compare. Without considering the contribution of charge carriers to thermal transport, the thermal conductivity $k$ is approximated as the lattice contribution $k_L$. For heavily doped samples, $k_L$ can be obtained by subtracting the computed electronic contribution $k_E$, as demonstrated in numerous studies.[55] In this work, experimental results are replaced with simulated $k_{R_b}$ and $k_{Ref}$ to demonstrate the data analysis. A total number of $M$ samples with varied neck width $w$ will be measured, keeping other geometric parameters unchanged. For the $i$th ($i = 1, 2, 3, \cdots M$) measurements, the ratio $r_i = k_{R_b}/k_{Ref}$ is given as

$$r_i = \int_0^\infty S(\eta_i) f(\Lambda_{eff}) d\Lambda_{eff} = \int_0^\infty K(\eta_i) q_i F(\Lambda_{eff}) d\Lambda_{eff}, \tag{1}$$

in which $\Lambda_{eff}$ is the effective in-plane phonon MFP within a solid thin film, $\eta = \Lambda_{eff}/w$, the Kernel function $K(\eta) = -dS/d\eta$, $q = d\eta/d\Lambda_{eff} = 1/w$, and $F(\Lambda_{eff}) = \int_0^{\Lambda_{eff}} f(\Lambda'_{eff}) d\Lambda'_{eff}$ is the accumulated phonon MFP distribution. The suppression function $S$ describes the in-plane heat flow reduction due to fabricated nanoslots. Integration by parts is used in the above conversion from $f(\Lambda_{eff})$ to $F(\Lambda_{eff})$, which also assumes $F(0) = 0$, $F(\infty) = 1$ and $S_i(\infty) = 0$. Here $\Lambda_{eff}$ is modified from the bulk phonon MFP $\Lambda_{Bulk}$ based on the film thickness $t$ and probability of specular phonon reflection (i.e., specularity) on film boundaries, known as the Fuchs-Sondheimer model.[30] Assuming completely diffusive film-surface phonon scattering, $\Lambda_{eff}$ is given as

$$\frac{\Lambda_{eff}}{\Lambda_{Bulk}} = 1 - \frac{3\Lambda_{Bulk}}{2t} \int_0^1 (x - x^3) \left[ 1 - \exp\left(-\frac{t}{\Lambda_{Bulk}} \frac{1}{x}\right) \right] dx. \tag{2}$$

The integral in Eq. (1) can be seen as an ill-posted question which could have infinite solutions. However, an unique solution of Eq. (1) can still be obtained without knowing the phonon transport information inside the structure.[56] In numerical integration, $F(\Lambda_{eff})$ at $N$ discretized $\Lambda_{eff}$ nodes is selected. Equation (1) is changed into

$$r_i = \sum_{j=1}^N K(q_i \Lambda_{eff,j}) q_i F(\Lambda_{eff,j}) \beta_j = \sum_{j=1}^N A_{i,j} F(\Lambda_{eff,j}), \tag{3}$$

where $\beta_j$ is the quadrature weight for each node, $q_i = 1/w_i$ for the $i$th sample, $A_{i,j} = K(q_i \Lambda_{eff,j}) q_i \beta_j$ is a $M \times N$ matrix. Here trapezoidal rule is used to determine quadrature points and weights.

Using convex optimization to match the measurement data for $M$ samples, the discretized MFP distribution $F(\Lambda_{eff,j})$ can be determined. The convex optimization aims to minimize a penalty function $P$:

$$P = \left\|AF - \frac{k_{Rb}}{k_{Ref}}\right\|_2^2 + \alpha\|\Delta^2 F\|_2^2. \tag{4}$$

In Eq. (4), $\Delta^2 F = F_{j+1} - 2F_j + F_{j-1}$ and $\|\cdot\|_2^2$ is the square of the second-norm. $k_{Rb}$ and $k_{Ref}$ represent the measured thermal conductance of the nanoslot-patterned thin film and solid thin film, respectively. In addition, $\alpha$ represents a smoothing factor to balance the accuracy penalty as the 1$^{st}$ term on the right hand side of Eq. (4), and the smoothness penalty as the 2$^{nd}$ term on the right hand side of Eq. (4). Some restrictions are imposed to this convex optimization: The accumulated phonon MFP distribution $F$ should be smooth and increase monotonically from $F(0) = 0$ to $F(\infty) = 1$. Detailed discussions of the convex optimization can be found in other studies,[13, 24, 57-59] in which a MATLAB-based package CVX was used for this purpose.[24, 56]

**2.3. Suppression function S predicted by phonon MC simulations and an analytical model**

One major task of the theoretical analysis is to obtain the employed $\eta$-dependent suppression function $S$. As one approach, $S(\eta)$ can be derived from the phonon BTE for some geometries.[60, 61] For completely ballistic phonon transport through an aperture such as the neck between nanoslots, simpler analytical models can be found.[49, 62] To extract $S(\eta)$ in Eq. (1), two-dimensional (2D) phonon transport is considered so that specular phonon reflection is enforced for the top and bottom surfaces of the film.[45] In this treatment, the influence of possibly diffusive phonon scattering by film surfaces is fully incorporated in $\Lambda_{eff}$ already and is not double counted in $S(\eta)$. On the other hand, rough nanoslot edges caused by nanofabrication should diffusively reflect phonons. Assuming a constant phonon MFP $\Lambda_{eff}$, thus a constant $\eta$ for one given $w$ value, frequency-independent phonon MC simulations are used here to predict $k_{R_b}$ for this given $\Lambda_{eff}$. The reference $k_{Ref}$ can be easily computed with the Fuchs-Sondheimer model[30] and numerous studies are available for Si thin films, leading to $S(\eta)=k_{R_b}/k_{Ref}$ for a particular set of $\Lambda_{eff}$ and $w$ values. The computed $S(\eta)$ is shown in Fig. 2a as crosses. The considered structure is a 1-μm-long film ($l = 1$ μm) with smooth film surfaces as the quasi-2D case. The neck width $w$ varies from 5 nm to 350 nm, with constant $b$=5 nm and $p$=500 nm.

Because frequency-independent phonon MC simulations can be time-consuming to acquire the cross symbols in Fig. 2a, an analytical model is also developed to estimate $S(\eta)$ as the lines in Fig. 2a. First, a characteristic length $L_C$ is assumed to modify the in-plane $\Lambda_{eff}$ due to two-dimensional patterns such as nanoslots and finite film length:

$$\frac{1}{\Lambda_{eff,0}} = \frac{1}{\Lambda_{eff}} + \frac{1}{L_C}. \tag{5}$$

In this case, the effective lattice thermal conductivity is calculated based on the kinetic relationship $k_L = cv_g\Lambda_{eff,0}/3$,[30] with the phonon specific heat $c$ and group velocity $v_g$ unchanged from bulk materials. For frequency-dependent analysis, $k_L$ is integrated over the whole phonon spectrum and summed up over different phonon branches. Then $k_{R_b}$ is computed as $k_{R_b} = kH_w \approx k_L H_w$, where $H_w$ is the correction factor to account for the reduced heat transfer in cross-sectional area due to nanoslots. This correction factor $H_w$ can be computed using the Fourier's law by comparing the thermal conductance of a nanoslot-patterned thin film and that of a solid thin film. Similar calculations can be found for periodic nanoporous Si films.[9, 63, 64] In this work, the thermal conductance calculation is performed using COMSOL Multiphysics software package.

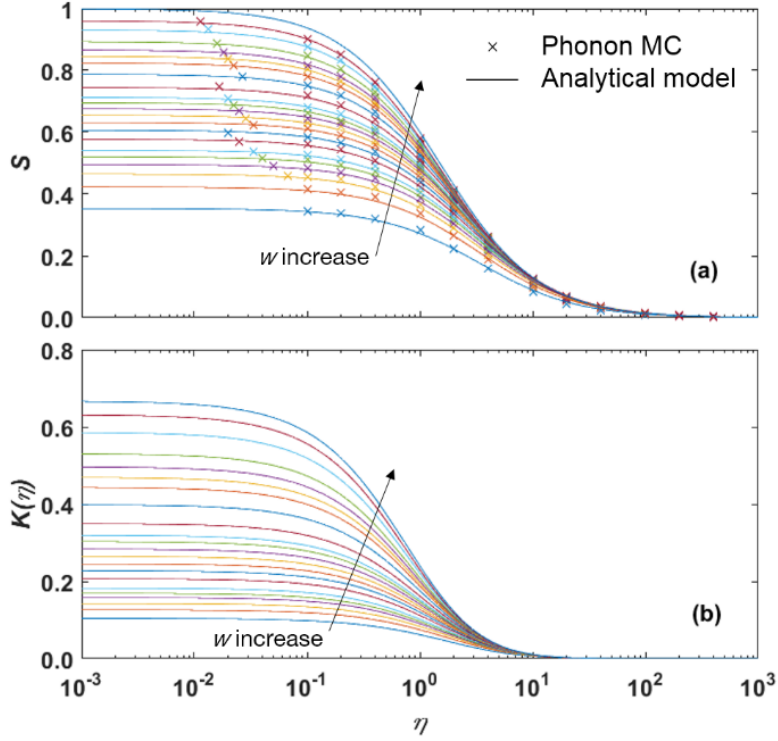

Figure 2. (a) Suppression function $S(\eta)$ Using Eq. (6) for quasi-2D cases. From the bottom to top curves, $w = 5, 10, 15, 20, 25, 30, 40, 50, 60, 70, 80, 90, 100, 120, 150, 180, 200, 220, 250, 300, 350$ and $500$ nm, respectively. Symbols are from phonon MC simulations assuming a constant phonon MFP. Symbols and curves have the same color for $w$ up to 350 nm. (b) Kernel function $K(\eta)$ using fitted $L_C$ values for $w=5-350$ nm and $L_C = 3l/4$ for $w= 500$ nm. Along the arrow direction, $w$ values match those in (a).

For the reference solid film, $k_{Ref}$ has $\Lambda_{eff}$ as the phonon MFP in the kinetic relationship. By comparing $k_{Ref}$ and $k_{R_b}$, an analytical model of the suppression function $S$ can be derived as

$$S(\eta) = \frac{\left(\frac{1}{\Lambda_{eff}}+\frac{1}{L_C}\right)^{-1}}{\Lambda_{eff}} H_w = \frac{L_C}{\Lambda_{eff}+L_C} H_w = \frac{L_C}{\eta w + L_C} H_w. \qquad (6)$$

When $\Lambda_{eff}$ approaches zero and leads to completely diffusive phonon transport, the Fourier's law is invoked so that $S(\eta = 0) = H_w$. At the limit $\Lambda_{eff} \to \infty$, $S(\infty)=0$ is obtained. With $H_w$ known for each $w$ values, one way to determine the corresponding characteristic length $L_C$ is to use Eq. (6) to fit the frequency-independent phonon MC simulation results, as curves in Fig. 2a. For $\eta < 0.01$, $S(\eta)$ saturates at $H_w$, i.e., the intersection between the $S(\eta)$ curve and the $x$-axis. With fitted $L_C$, the Kernel function $K(\eta)$ can be acquired analytically (Fig. 2b). As an extreme case with $w = p = 500$ nm and thus no nanoslots, $L_C = 3l/4$ is directly used to compute $S(\eta)$ for a solid film[65] and fitting with frequency-independent phonon MC simulations is unnecessary.

Alternatively, $L_C$ can also be derived by examining the thermal conductance of the nanoslot-patterned thin film at the limit $\Lambda_{eff} \to \infty$. At this limit, the neck between two adjacent nanoslots functions as an aperture for the phonon transport. In this situation, the neck receives black-body emission from the thermal reservoirs at both ends of the structure (Fig. 1a). Following a similar derivation as in previous studies,[62] the associated thermal conductance $G_{Rb}$ for one period is

$$G_{Rb} = \frac{Cv_g wt}{4}, \qquad (7)$$

where the product $wt$ stands for the cross-section area of the neck. This $G_{Rb}$ is derived as the net heat flow across a neck divided by the temperature difference applied across the whole structure. More associated analysis can be found elsewhere.[49, 66] Along another line, Eq. (5) suggests $\Lambda_{eff,0} = L_C$ at $\Lambda_{eff} \to \infty$ so that

$$G_{Rb} = \frac{kH_w pt}{l} = \frac{\left[\frac{1}{3}Cv_g L_C\right]H_w pt}{l} = \frac{Cv_g L_C H_w pt}{3l}. \qquad (8)$$

Comparison between Eqs. (7) and (8) yields an analytical expression of $L_c$:

$$L_c = \frac{w}{p}\frac{3l}{4H_w}, \qquad (9)$$

which can also be used in Eq. (6) for $S(\eta)$ predictions. In Eq. (9), the structure simply becomes a solid film at $w = p$. In this extreme case, the correction factor $H_w = 1$ and the corresponding $L_c = 3l/4$ in Eq. (9). This $L_c$ value also matches that proposed for a solid film in an early study.[65] Figure 3 shows the comparison between fitted and derived $L_c$ values. The divergence is within 5% for most $w$ values. To probe the contribution of phonons with down to ~10 nm MFPs, here $L_c$ is down to 16.9 nm for the smallest $w$=5 nm. Despite a reasonably-well agreement between the fitted and derived $L_c$ values, particular attention should be paid to the fitted $L_c$ value at $w$ <10 nm (Solid line in Fig. 3 inset), which becomes smaller than predictions by Eq. (9). This is due to an additional conduction thermal resistance $R_{neck} = b/(wtk_{neck})$ added by the neck region with depth $b$, where $k_{neck}$ is the effective thermal conductivity for the neck region with diffusive phonon boundary scattering on its sidewalls. By reducing $b$ to zero or setting specular phonon reflection for the neck sidewalls, the divergence between the fitted and predicted $L_c$ values can be eliminated.

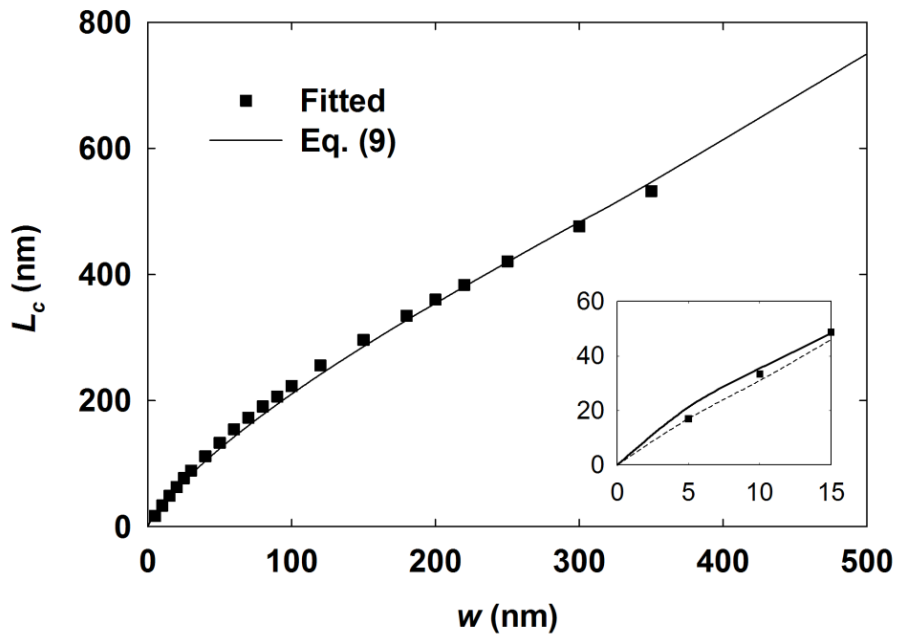

Figure 3. Comparison between derived and fitted $L_c$.

To further investigate the influence of $b$ on $L_c$, 2D structures with fixed $w, l, p$ but varied $b$ are simulated. In principle, the neck region is in analogy to a thin film with its film thickness as the neck width $w$ here. Assuming completely diffusive film-surface phonon scattering as that for the neck sidewalls, Eq.(2) can be used to modify $\Lambda_{Bulk}$ for the effective in-plane MFP within the neck region $\Lambda_{neck}$, yielding $k_{neck}$ and $R_{neck}$. For the whole film, this $R_{neck}$ is added to the predicted $1/G_{Rb}$ for a nanoslot-patterned film with the neck depth $b \to 0$. The solid lines in Fig. 4 present the predicted total thermal resistance $R_{total}$ of different 2D nanoslot-patterned films with a fixed thickness of 10 nm. Specular phonon reflection is assumed on the top and bottom surfaces of these 2D structures. The possible change of the phonon dispersion and phonon scattering rates within the neck region is neglected here but can be included in more accurate analysis, as shown in phonon studies of nanowires.[67-69] The prediction agrees well with phonon MC simulations (symbols in Fig. 4). In practice, $b$ should be minimized in nanofabrication. Compared to ballistic thermal resistance $R_b$, $R_{neck}$ is thus negligible and Eq. (9) can be used to accurately determine $L_c$. When $b$ is relatively large, Eq. (7) becomes

$$\frac{1}{G_{total}} = \frac{4}{Cv_g wt} + \frac{b}{wtk_{neck}} = \frac{1}{Cv_g wt}\left(4 + \frac{3b}{\Lambda_{neck}}\right). \tag{10}$$

Comparing Eq. (10) with Eq. (8) yields

$$L_c = \frac{w}{p}\frac{3l}{(4+3b/\Lambda_{neck})H_w}. \tag{11}$$

This modified $L_c$ is plotted in the inset of Fig. 3 as a dashed line and shows better agreement with the phonon MC simulations.

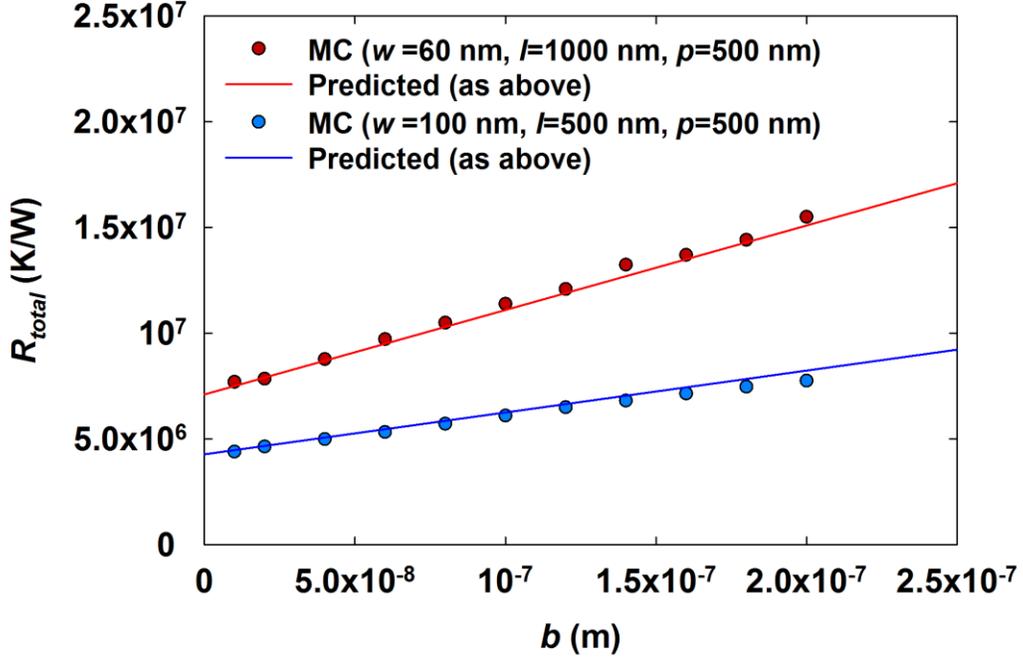

Figure 4.  Comparison between the predicted and MC-simulated $R_{total}$.

## 3. RESULTS AND DISCUSSION

### 3.1. Temperature profiles within nanoslot-patterned Si films sandwiched between two blackbodies

As virtual experiments, the room-temperature $k_{R_b}$ and thus $k_{R_b}/k_{Ref}$ ratio are first computed by frequency-dependent phonon MC simulations for the proposed thin film with varied neck width $w$ between nanoslots.  Without losing the generality, the first-principles phonon MFPs for bulk Si,[7] as validated by experiments,[13] are used in phonon MC simulations.  In phonon MC simulations, a film thickness $t$ of 70 nm is assumed, with rough film surfaces for completely diffusive phonon reflection.  Other geometry parameters are consistent with those used in Fig. 2, i.e., $l = 1$ μm, $b = 5$ nm, and $p = 500$ nm.

Figures 5(a)–(d) show the temperature contours of representative nanoslot-patterned thin films with two ends in contact with blackbody heat reservoirs fixed as 305 and 295 K, respectively.

For small $w$ values, particular attention should be paid to the long time required for the phonon MC code to converge. In principle, the chance for a phonon to pass the neck regions becomes very small with a decreased $w/p$ ratio. Therefore, a longer simulation time is needed to statistically obtain the impact of the neck region on the thermal conductivity of the whole film. Figure 5(e) further presents the heat-flow-direction temperature profiles along the symmetric line of a period.

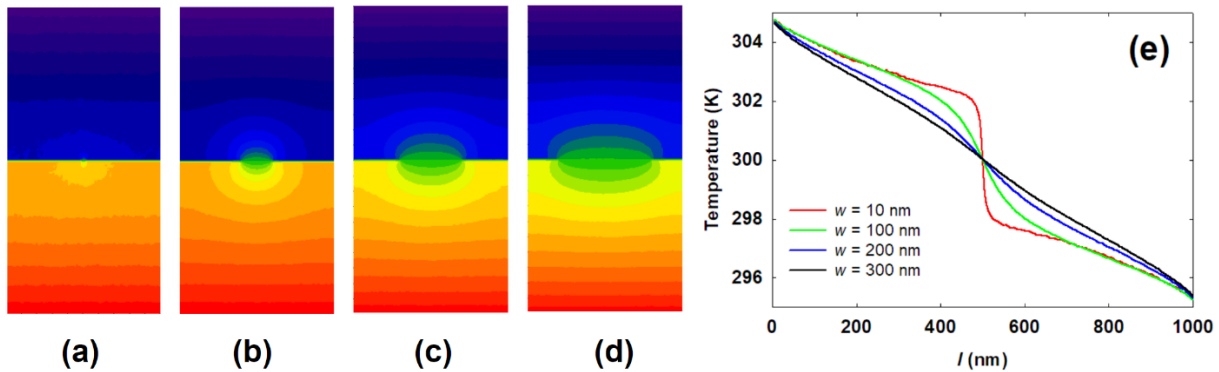

Figure 5. Temperature contours of a nanoslot-patterned thin film with $l = 1$ µm, $b = 5$ nm, $p = 500$ nm. The neck width $w$ is (a) 10 nm, (b) 100 nm, (c) 200 nm, and (d) 300 nm, respectively. (e) Temperature profiles along the central symmetric line of a period in (a) to (d).

### 3.2. Reconstructed phonon MFP distribution

The $k_{R_b}/k_{Ref}$ ratios given by frequent-dependent phonon MC simulations are shown as a function of $w$ in Fig. 6a. The values of $w$ match those in Fig. 2. As anticipated, the $k_{R_b}/k_{Ref}$ ratio monotonously increases as $w$ is expanded to the pitch $p$. However, this ratio cannot reach 100% at $w = p$ due to the phonon MFP constriction by the finite film length $l$, where $3l/4$ should be used to modify the bulk phonon MFP in $k_L$ predictions.[65] Following the description in Section 2.2, $F(\Lambda_{eff})$ can be reconstructed through convex optimization. The fitted $L_c$ is used to take the depth $b$ influence into consideration. Here $M = 21$ samples and $N = 200$ logarithmic-spaced discretized intervals are used in the MFP reconstruction. The obtained MFP distribution function $F$ is found to be less sensitive to the smooth factor $\alpha$ in the range of $0.5 < \alpha < 2.0$, as indicated by the colored band in Fig. 6(b). A smooth factor $\alpha = 1.0$ is adopted. The reconstructed $F(\Lambda_{eff})$ generally agrees with the employed first-principles $F(\Lambda_{eff})$. Some divergence is anticipated because the reconstructed $F(\Lambda_{eff})$ is a smooth curve, in contrast with the irregular first-principles $F(\Lambda_{eff})$ curve. For different $\alpha$ values, however, the reconstructed $F$ always matches well with employed first-principles $F$ at $F \approx 60\%$. At the smallest $w$=5 nm, the characteristic length is $L_c$=16.9 nm, which could limit the accuracy of the phonon MFP reconstruction in the sub-10 nm regime. To check this, the same inverse phonon transport studies are carried out after adding one more $k_{R_b}/k_{Ref}$ value at $w$= 1 nm, with the corresponding characteristic length reduced to $L_c$=5.0 nm. Repeating the same procedure with $\alpha$=1.0, the newly reconstructed $F(\Lambda_{eff})$ is almost identical to $F(\Lambda_{eff})$ using a minimum $w$= 5.0 nm, as shown in Fig. 6(c). This negligible difference is also due to the relatively weak thermal conductivity contribution by phonons with sub-10 nm MFPs. Some divergence is anticipated when these ultra-short-MFP phonons become important to heat conduction, e.g., samples with significantly reduced phonon MFPs at high temperatures.

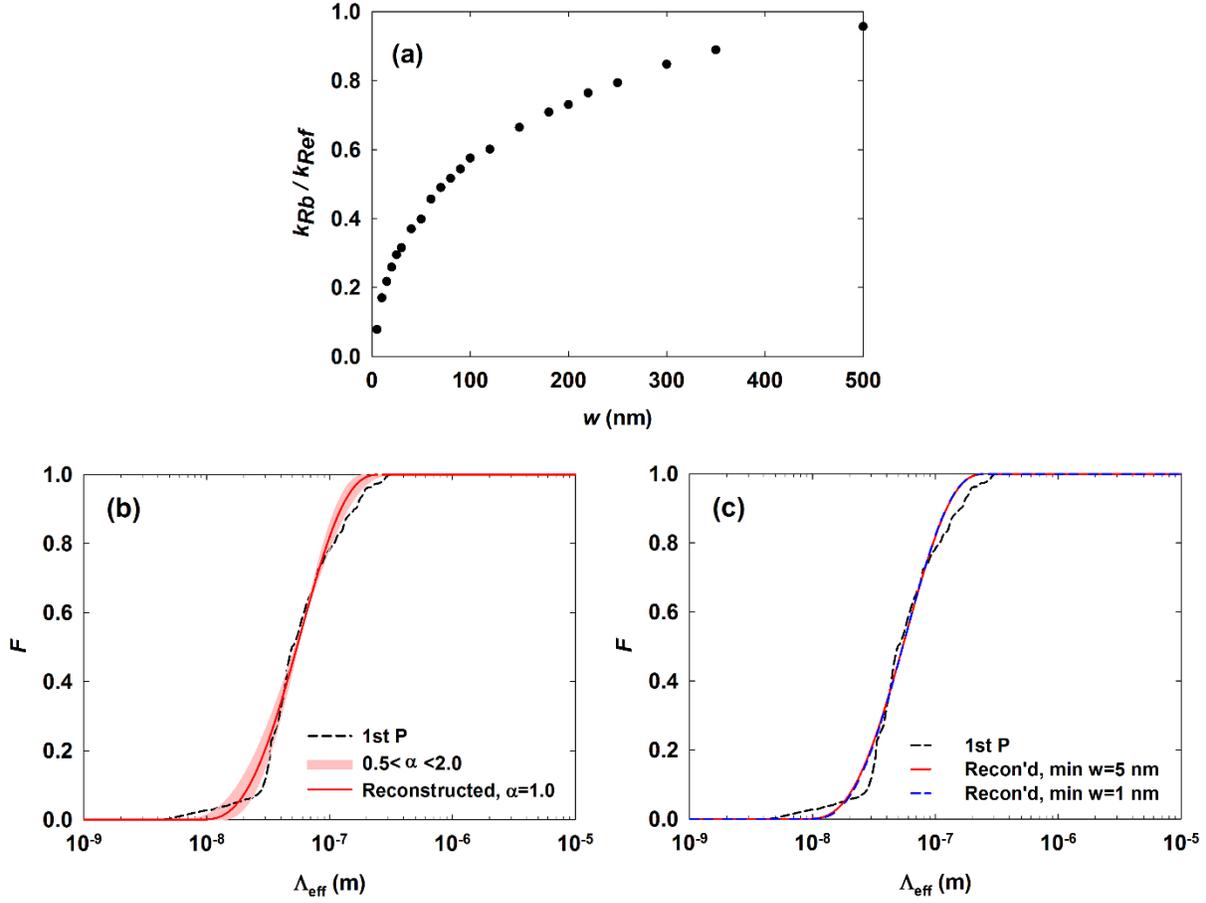

Figure 6. (a) $k_{R_b}/k_{Ref}$ ratio computed with first-principles bulk Si phonon MFPs in frequency-dependent phonon MC simulations. (b) Comparison between reconstructed MFP distribution $F(\Lambda_{eff})$ and the employed first-principles $F(\Lambda_{eff})$. (c) Three-way comparison between the employed fist-principles $F(\Lambda_{eff})$, reconstructed MFP distributions with a minimum $w$=5 nm or $w$=1 nm.

### 3.3. Tuning the characteristic length to better probe short-MFP phonons

To better sense ultra-short phonon MFPs, a smaller minimum $L_c$ is required. According to Eq. (9), $L_c$ is not only restricted by $w$ but also related to $l$ and $p$. Figures 7(a) and 7(b) present the variation of $H_w$ and $L_c$ with respect to $w$ and $l$, respectively. Here $p$=500 nm and $b$=1 nm are fixed. In Fig. 7(a), it can be observed that $H_w$ becomes larger as $l$ increases, which is due to the increased importance for the length-direction thermal resistance added to the ballistic thermal resistance $R_b$. Figure 7(b) presents the $L_c$ distribution with respect to $w$ and $l$, where $L_c$ has a stronger dependence on $w$. The lowest $L_c$=10.2 nm is found at the minimum $w$ =5 nm and $l$= 100 nm. In nanofabrication, $w$<5 nm cannot be easily achieved with the standard electron beam lithography and dry etching. For suspended 2D materials, however, such a small neck region can be directly patterned with electron beam drilling.[70]

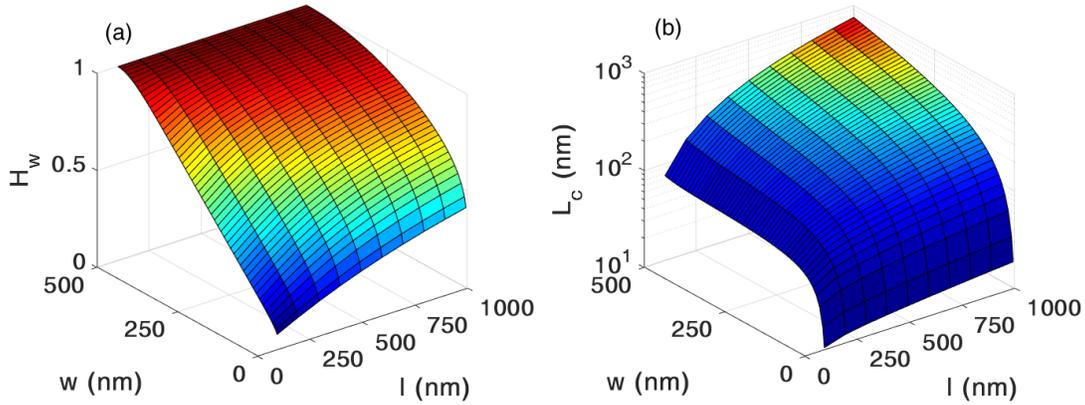

Figure 7. The dependence of (a) reduction factor $H_w$ and (b) $L_c$ on $w$ and $l$. Other dimensions are fixed at $p$=500 nm and $b$=1 nm.

### 3.4. Comparison between periodic nanoslot and nanopore patterns

The proposed nanoslot-patterned thin films form an interesting comparison with widely studied periodic nanoporous Si films.[31-37] It is now acknowledged that coherent phonon transport is only critical at low temperatures, where the dominant phonon wavelength scales up with $1/T$ and becomes comparable to the structure sizes. By comparing the thermal conductivities of periodic and aperiodic nanoporous film, wave effects due to coherent phonon interference was identified below 14 K for Si films with $p$>100 nm,[36] or below 10 K for $p$=300 nm.[31] In most cases, incoherent phonon transport is dominant for the thermal conductivity reduction, which is similar to nanoslot-patterned thin films.

For the phonon MFP reconstruction, one major problem for using such periodic nanoporous films is the lack of an accurate analytical model based on the phonon MFP modification using an effective characteristic length $L_C$. In principles, using the Matthiessen's rule in Eq. (5) to combine boundary phonon scattering and internal phonon scattering within a volume is not accurate.[30] For aligned nanopores with a pitch $p$ and pore diameter $d$, MC ray tracing suggests $L_C = (4p^2 - \pi d^2)/\pi d$ as the geometric mean beam length (MBL) for the structure.[9, 71] This $L_C$ is accurate when ballistic phonon transport is dominant in the structure. However, it is less accurate with increased internal phonon scattering and some modifications of $L_C$ should be taken.[9, 44] In the literature, the proposed characteristic lengths also include $L_C = \sqrt{p^2 - \pi d^2/4}$ as the square root of the solid area within a period,[41] $L_C = p - d$ as the neck width,[39, 40] $L_C \sim 4p^2/\pi d$ as the reciprocal of the surface-to-volume ratio.[38] None of these $L_C$ values are consistently accurate for arbitrary pitch-diameter combinations.[9, 44] Clearly $L_C = \sqrt{p^2 - \pi d^2/4}$ and $L_C = p - d$ are only valid for dense nanopores because $L_C > p$ is anticipated for sparse nanopores patterned across a film. In general, a consistent $L_C$ cannot be expected for Eq. (5) over the whole phonon MFP spectrum. The lack of an accurate analytical model for the predictions of the suppression function $S(\eta)$ and the kernel function $K(\eta)$ can largely increase the difficult level of the inverse phonon transport analysis. In practice, $S(\eta)$ and $K(\eta)$ can be obtained with a MC technique based on ray-tracing or path sampling,[36, 72, 73] or by solving the MFP-dependent phonon BTE.[74] However, these techniques are very complicated and cannot be used to predict the whole $S(\eta)$ and $K(\eta)$ curves in an effective way. For a nanoslot-patterned thin film, phonons are mainly affected by the ballistic thermal resistance introduced by the neck width $w$. The structure complexity is largely reduced from that for periodic nanoporous Si films with varied pitches and pore diameters.

Within periodic nanoporous films, the shape variation of individual nanofabricated pores across the whole film can add large uncertainties to the data analysis, along with varied pore-edge defects or roughness to affect the effective pore diameter and thus the neck width.[43, 46-48] This makes such periodic nanoporous Si films unsuitable for the study of the phonon MFP distribution. For a thin films with only a few nanoslots, above uncertainties can be largely minimized in the data analysis using an averaged $w$ revealed by the electron microscopy. The effective $w$ value for each neck can be slightly expanded due to the amorphous pore edges, as suggested for nanoporous Si films in electron microscopy studies.[34, 46, 47]

## 4. CONCLUSIONS

In summary, a new approach to determine the phonon MFP distribution $F(\Lambda_{eff})$ for the in-plane heat conduction in thin films is proposed in this work. An analytical solution for the suppression function $S$ of nanoslot-patterned thin films is also proposed and can be employed to compute the transport properties of two-dimensional materials using such patterns. For thin films and 2D materials, their in-plane phonon MFP distribution can be obtained with the proposed approach, whereas the widely-used pump probe measurements can be largely restricted for suspended structures. Although the relaxation time approximation is found to be less accurate for 2D materials due to the importance of normal processes,[75] such phonon MFP reconstructions can still yield the range of phonons MFPs[76] to be compared with existing calculations.[77]

Beyond phonon MFP reconstructions, the proposed nanoslot patterns can also be used to enhance the thermoelectric performance of general thin films, in addition to the widely studied periodic nanoporous patterns.[31-37] When the characteristic length $L_C$ is shorter than majority phonon MFPs but still longer than the dominant electron MFPs, the thermal conductivity can be dramatically reduced but the electrical properties can still be maintained, leading to improved thermoelectric performance. In the analysis based on the BTE, the proposed $L_C$ expression can be directly used to modify the electron and phonon MFPs to compute the thermoelectric properties.

## ACKNOWLEDGMENTS

The authors acknowledge the support from the U.S. Air Force Office of Scientific Research (award number FA9550-16-1-0025) for studies on nanoporous materials and National Science Foundation (grant number CBET-1651840) for phonon simulations. An allocation of computer time from the UA Research Computing High Performance Computing (HPC) and High Throughput Computing (HTC) at the University of Arizona is gratefully acknowledged. The authors would also like to thank Prof. Austin Minnich and Dr. Emigdio Chávez-Ángel for the valuable insights on convex optimizations, and Dr. Hongbo Zhao for the help on phonon MC simulations.